\documentclass[manuscript, sigconf]{acmart}

\AtBeginDocument{%
  \providecommand\BibTeX{{%
    \normalfont B\kern-0.5em{\scshape i\kern-0.25em b}\kern-0.8em\TeX}}}

\usepackage{hyperref} 

\usepackage{hyperxmp} 


\copyrightyear{2026} 
\acmYear{2026} 
\acmConference[SIGCSE '26]{Proceedings of the 56th ACM Technical Symposium on Computer Science Education}{Feb 18--21, 2026}{St. Louis, MO, USA}

\begin{document}

\title[Designing for Learning with GenAI is a Wicked Problem]{Designing for Learning with Generative AI is a Wicked Problem: An Illustrative Longitudinal Qualitative Case Series}

\author{Clara Scalzer}
\email{maria.scalzer@maine.edu}
\orcid{}
\affiliation{%
  \institution{University of Maine}
  \streetaddress{}
  \city{Orono}
  \state{Maine}
  \country{USA}
  \postcode{xx}
}
\author{Saurav Pokhrel}
\email{saurav.pokhrel@maine.edu}
\orcid{}
\affiliation{%
  \institution{University of Maine}
  \streetaddress{}
  \city{Orono}
  \state{Maine}
  \country{USA}
  \postcode{xx}
}
\author{Sara Hunt}
\email{sara.hunt@maine.edu}
\orcid{}
\affiliation{%
  \institution{University of Maine}
  \streetaddress{}
  \city{Orono}
  \state{Maine}
  \country{USA}
  \postcode{xx}
}
\author{Greg L Nelson}
\email{gregory.nelson@maine.edu}
\orcid{}
\affiliation{%
  \institution{University of Maine}
  \streetaddress{}
  \city{Orono}
  \state{Maine}
  \country{USA}
  \postcode{xx}
}

\begin{abstract}

Students continue their education when they feel their learning is meaningful and relevant for their future careers. Computing educators now face the challenge of preparing students for careers increasingly shaped by GenAI with the goals of supporting their learning, motivation, ethics, and career development. Our longitudinal qualitative study of students in an GenAI-integrated creative media course show how this is a “wicked” problem: progress on one goal can then impede progress on other goals. Students developed concerning patterns despite extensive instruction in critical and ethical GenAI use including prompt engineering, ethics and bias, and industry panels on GenAI's career impact. We present an analysis of two students’ experiences to showcase this complexity. Increasing GenAI use skills can lower ethics; for example, Pat started from purposefully avoiding GenAI use, to dependency. He described himself as a "notorious cheater" who now uses GenAI to "get all the right answers" while acknowledging he's learning less. Increasing ethical awareness can lower the learning of GenAI use skills; for example Jay’s newfound environmental concerns led to self-imposed usage limits that impeded skill development, and new serious fears that GenAI would eliminate creative careers they had been passionate about. Increased GenAI proficiency, a potential career skill, did not improve their career confidence.
These findings suggest that supporting student development in the GenAI era is a “wicked” problem requiring multi-dimensional evaluation and design, rather than optimizing learning, GenAI skills, ethics, or career motivation individually.

\end{abstract}


\maketitle

\section{Introduction}

    Generative AI (GenAI) is rapidly reshaping the educational landscape. A growing body of research highlights GenAI's potential to enhance personalized learning, foster creative problem-solving, and streamline academic processes, making educational experiences more engaging and efficient. Many students share this optimism, viewing GenAI as a time-saving tool, focus on more strategic work, and ultimately advance their careers \cite{chan_students_2023}. From this perspective, integration is a “tame” problem: complex, but ultimately solvable with the right tools and pedagogical strategies.

    However, this paper argues that the challenge is far more profound, and what we need to evaluate success is more complex than it appears. We contend that designing for learning with GenAI is a "wicked problem"—a term for problems with no clear solution, marked by entangled, often unintended consequences \cite{RittelDilemmasTheory1973}. Unlike a tame problem with clear inputs and outputs, wicked problems require trade-offs across competing values, and addressing one goal (like GenAI skills) may worsen others (like ethical behavior or career confidence). Our data suggests that GenAI-integrated education has these characteristics today.
    
    The challenge is not just that GenAI might fail to produce correct answers, but that it often “succeeds” in ways that undercut the learning process itself. A student may generate a passing assignment in seconds—without struggle, reflection, or human feedback. That efficiency can reinforce shortcuts over understanding, isolating students from peers and instructors and weakening social bonds that support learning and motivation \cite{hou_all_2025,crawford_when_2024}. Weakened social connection can, in turn, diminish a student’s sense of belonging, reduce motivation, and increase reliance on the isolating efficiency of GenAI—patterns that recent work is beginning to document in computing \cite{hou_all_2025}, but there are few longitudinal studies that might detect such change over time.
    
    This destructive cycle can be interpreted through Social Cognitive Career Theory (SCCT), which explains how career interests form through interactions between task performance, self-efficacy, and outcome expectations \cite{lentSocialCognitiveModel2013}. When students believe they succeed due to their own effort, self-efficacy grows—fueling motivation and career confidence. However, GenAI can short-circuit this loop: it may enable high performance without deep engagement, creating what Prather et al. (2024) call an “illusion of competence”—a mismatch between perceived and actual ability that can erode resilience when students face complex tasks \cite{prather2024widening}. While the illusion may feel like genuine competence in the short term, it may not prepare students to adapt or recover when the scaffolding of GenAI is absent. This fragility becomes more concerning in light of findings by Margulieux et al., who report complex relationships between GenAI use and self-efficacy among novice programmers—including reduced confidence or increased fear of failure after relying on GenAI tools \cite{margulieux_self-regulation_2024}. SCCT also emphasizes how perceptions of the labor market influence career motivation: if students believe GenAI threatens their future profession, they may lower their expectations—even while achieving high marks \cite{sowaSupportingChildrensCareer2023}. These intersecting effects exemplify the wicked problem framing, where interventions aimed at improving one area—like learning outcomes—can unintentionally destabilize others, such as ethical reasoning or career confidence \cite{RittelDilemmasTheory1973}.

    Therefore, educators can no longer afford to ask only, "Did this intervention improve learning on this task?" They must also ask, "What did this intervention do to the student's motivation, ethical reasoning, confidence in their future, and their connection to peers and discipline?" A single cross-sectional snapshot cannot capture the cumulative effects and feedback loops involved. Only by tracing students over time can we observe the complex interplay—and sometimes unintended consequences—of GenAI use on learning, ethics, and career development.

    This paper offers the first longitudinal, in-depth qualitative study of how students experience GenAI in an educational setting explicitly designed to support ethical reasoning, GenAI skill-building, and career awareness. Our analysis of two richly documented student trajectories reveals how these goals can come into conflict: progress on one dimension can impede another. We do not argue that GenAI integration in education must always be a wicked problem, but our findings demonstrate that in this setting, it exhibited clear characteristics of one: unpredictable outcomes, competing values, and evolving dilemmas without straightforward resolution.

    We observed these wicked dynamics while answering the following research questions:

\begin{itemize}
    \item \textbf{RQ1}: How do the effects of using GenAI change over time?
    \item \textbf{RQ2}: How do students’ perceptions of GenAI’s impact on their careers evolve over time?
    \item \textbf{RQ3}: How do students’ expectations about their own career outcomes change over time?
\end{itemize}

\section{Related Work}

We ground our analysis in the literature on wicked problems, which are defined as ill-structured, unsolvable in definitive terms, and symptomatic of other entangled challenges \cite{RittelDilemmasTheory1973}. Unlike “tame” problems, wicked problems lack clear endpoints and often generate new dilemmas when addressed. Educational researchers have described many teaching challenges—such as designing inclusive curricula, supporting diverse learning needs, or teaching ethical technology use—as wicked problems due to their persistent, value-laden trade-offs \cite{bass_coda_2022, roberts_wicked_2000}. Integrating emerging technologies like generative AI into pedagogy adds another layer of complexity, as students' cognitive, ethical, and career development needs often pull in conflicting directions.

Wicked problems have long been a framework for understanding complex challenges in education, where competing goals, diverse stakeholders, and unpredictable outcomes resist linear solutions \cite{churchman_guest_1967, RittelDilemmasTheory1973}. In educational contexts, these challenges include curriculum reform, systemic inequity, and technology integration—each marked by incomplete knowledge and value-laden tradeoffs \cite{camillus_strategy_2008, kolko_wicked_2012}. Kolko, for instance, argues that curriculum should intentionally embrace ambiguity, enabling learners to navigate competing perspectives rather than resolve them \cite{kolko_wicked_2012}. In computing education, Mishra and Koehler characterize the integration of technology into teaching as a wicked problem itself—requiring complex pedagogical decisions and constant adaptation \cite{mishra_technological_2007}. These perspectives align with broader educational scholarship advocating for participatory, reflective, and value-sensitive approaches to instruction and reform \cite{roberts_wicked_2000, bass_coda_2022}. Our study extends this tradition by examining how students experience generative AI not just as a tool but as a catalyst for ethical, motivational, and identity-based dilemmas, thereby highlighting the deeply wicked nature of integrating such technologies into learning environments.

While the wickedness of educational innovation has been acknowledged conceptually, few studies have empirically examined how these tensions manifest over time in students' real experiences. To our knowledge, no prior work has longitudinally traced how students’ engagement with GenAI evolves while examining intersecting outcomes such as learning, ethics, motivation, and career outlook. Dai et al. \cite{dai_why_2025} conducted a valuable multi-timepoint study, yet it focused on design motivations within engineering education and did not track individual students' ethical or career reasoning; similar studies include \cite{keuning_students_2024, boguslawski_programming_2025, zastudil_generative_2023}. Studies \cite{kuhail_will_2024} like Yasar and Karagücük’s \cite{yasar_effect_2025} examine anxiety and GenAI career threats quantitatively, while Sallam et al. \cite{sallam_anxiety_2024} compare career outlooks across disciplines. These provide broad insights but lack the depth of longitudinal, qualitative tracing of student transformation.

\subsection{Empirical Studies on Student GenAI Use}

A growing body of work has explored student interactions with generative AI tools. Survey-based studies capture general attitudes \cite{chan_deconstructing_2023, sallam_anxiety_2024}, tool adoption rates, and perceived risks or benefits \cite{yasar_effect_2025, black_university_2025}. Some report that students adopt GenAI when its utility outweighs ethical concerns \cite{chan_deconstructing_2023}, while others show how tool use amplifies or dampens creative self-efficacy \cite{hwang_influence_2024}. However, these studies often isolate a single outcome—creativity, anxiety, or skill development—without examining how those outcomes intersect over time.

Qualitative research has provided more context-rich portraits of GenAI use in learning \cite{prather_beyond_2025, zastudil_generative_2023,hou_all_2025, hou_effects_2024}. Black and Tomlinson \cite{black_university_2025}, for example, found that students in a writing-intensive course used GenAI cautiously and often limited it to proofreading, citing concerns about authorship and intellectual independence. These studies offer useful snapshots; however, they are short-term in scope and do not trace students’ changing perceptions or behaviors over time.

While Dai et al. \cite{dai_why_2025} collected design journals across a course, their focus remained on design motivation, and not on ethics, learning, or career identity. No study to date has combined rich qualitative data with wicked problem theory to examine longitudinal trajectories of GenAI engagement in ethics-integrated, creative coursework. Our study fills this gap by tracing student transformations across time and across multiple domains of development.

In addition to wicked problems theory, we draw on Social Cognitive Career Theory (SCCT) to interpret how students’ beliefs about their GenAI skills, ethical agency, and expected career outcomes evolved. SCCT emphasizes self-efficacy, outcome expectations, and goal setting as drivers of career decision-making. The challenge of these psychological drivers is amplified by the wicked nature of GenAI integration, where problems are ill-defined, constantly changing, and laden with ethical tradeoffs. Prior survey work shows that GenAI-related anxiety correlates with lower career certainty, especially in disciplines like translation, where GenAI poses a direct automation threat \cite{yasar_effect_2025}. Meanwhile, tool familiarity can improve confidence and broaden perceived career possibilities \cite{hwang_influence_2024}. However, the feedback loops proposed by SCCT—where self-efficacy, experience, and outcome expectations reinforce one another—have rarely been studied qualitatively in GenAI-integrated coursework. Our paper demonstrates how these dynamics unfold in real time, often in tension with ethical concerns and institutional norms.

\section{Method}

We conducted a longitudinal qualitative study to examine how students’ perceptions of GenAI and its role in their careers evolved over one semester. Our methods were guided by research questions focused on change over time in an undergraduate creative media course. The study received Institutional Review Board (IRB) approval from the Anonymous University.

\label{methods:leanrning-context}

The study took place in a Fall 2024 introductory creative media course at [Anonymized University], a public rural research university in the northeastern U.S. The curriculum introduced novice creators to a range of digital tools, including traditional and GenAI software for writing, image manipulation, programming, and sound. The course's weekly interleaving had students alternate between completing a task with conventional tools and repeating it using GenAI. Each GenAI task was paired with instruction on prompt engineering and followed by ethical reflection. The course was deliberately structured to integrate GenAI in a pedagogically meaningful way. Students received explicit instruction on prompting techniques, ethical reasoning, and evaluating GenAI-generated outputs. Class discussions and written reflections further encouraged students to examine the broader implications of using these tools.

We recruited students through an in-class announcement and email. While 19 expressed interest, 14 of 54 enrolled students completed the full study. This paper presents an in-depth analysis of two participants to illustrate a range of experiences. Pat is a 21-year-old White man in his fourth year with a disability. Jay is a 21-year-old White trans-masculine student, also in their fourth year and disabled. Both identified as novices with GenAI and creative tools. We selected Pat and Jay because their interviews and reflections provided the most detailed and thought-provoking accounts of how students navigated GenAI use over time.

Our longitudinal approach had four semi-structured interviews with each of the 14 participants, then held approximately once per month throughout the semester. The first interview was done before GenAI was introduced in the course (before week 3). The semi-structured interview protocol included questions about their evolving opinions on GenAI, its use in and out of the class, their ethical guidelines, their career interests, perceived future careers, and the perceived impact of GenAI on their future professions. However, the analysis for this paper focuses specifically on the first and final interviews (interview 1 and interview 4). This approach was chosen to provide a direct comparison of students' perspectives at the beginning and end of the course, aligning with our research questions investigating change over time. Each interview lasted between 42 and 102 minutes (mean 64) and was conducted remotely via Zoom. Zoom interviews allowed for recording, scheduling flexibility, and cross-institution collaboration.

\subsection{Analysis}

All interviews were transcribed and human-corrected verbatim for analysis and review. To answer our research questions, our analysis team performed a thematic analysis using Atlas.ti without AI. 

The research team developed a codebook based on our research questions and SCCT theory. The complete transcripts from the first and final interviews for the two focus participants were analyzed using this framework. A team of three researchers coded the data and met regularly to discuss interpretations, refine codes, and reach agreement. Rather than calculating inter-rater reliability, we prioritized resolving disagreements through collaborative discussion to deepen shared understanding, consistent with qualitative research practices in computing education \cite{salac_funds_2023, everson_key_2022, hammer_confusing_2014} and HCI \cite{mcdonald_reliability_2019}. Each researcher drafted a pen portrait, which is a narrative synthesis capturing a participant’s experience over time, following Sheard and Marsh’s four-stage method: define focus, design structure, populate content, and interpret findings \cite{sheard_portrait_2019}. These portraits addressed each research question and highlighted observed changes. The team then discussed and reached consensus before writing final results.

\section{Results}

\subsection{Pat - Raised AI skills, Lowered ethical skills}

Pat is a senior business management major with a concentration in entrepreneurship from a small coastal town in [northeastern state]. In interview 1, he described himself as a "notorious cheater" while also refusing to use GenAI in his school work. He initially held a negative view of GenAI, viewing it as "trash", but his experiences in the creative media course led him to accept and use it for academic efficiency. His perception of GenAI's career impact remained consistent, believing it would take jobs but not the ones he wanted, and his career outcome expectancy remained consistent.

\subsubsection{Change in effects of using GenAI over time}

In the first interview, Pat was highly skeptical of GenAI, having had no experience with it before this semester. He viewed it as "trash" and untrustworthy for academic work like essays, fearing plagiarism. As a self-described "notorious cheater", he actively avoided GenAI because "it seemed like a trap", and he knew that if he started, "\textit{I'm never gonna not use it}". However, he had recently started using Google's GenAI search feature for homework questions and was impressed by its ability to "pull out the answer immediately", which he found made his work easier.

In the fourth interview, Pat had completely changed his stance, now viewing GenAI "\textit{as a tool now that I can use to help me do schoolwork}". His confidence grew "pretty quickly" due to in-class work and observing his professor's effective prompting techniques. Pat uses his own discretion for what he deems valuable to learn- he admitted to using GenAI to get perfect grades on "meaningless multiple choice questions" for "stupid class(es)", stating, "\textit{if I'm going to submit it for a grade, I'm definitely going to use AI to get all the right answers}". He values its efficiency above all, believing that "AI just does things faster and better." He still held the view that using GenAI to write an essay "cheats" the learning process; however, he was currently using GenAI to "\textit{go over essays, correct essays, add stuff to essays}". Pat stated, "\textit{I really didn't have any ethical views before, and I still don't really}".

Pat's perspective on the effects of using GenAI changed dramatically, from an active avoidance based on fear of getting caught and poor output quality, to a full adoption of GenAI for academic efficiency. His initial skepticism, rooted in the belief that he could cheat better than GenAI, evolved into a mindset where the most efficient method was the best method. For Pat, GenAI became a tool to "get all the right answers" with less effort, particularly for academic work he deemed unimportant. This quick adoption was fueled by an appreciation for efficiency and no ethical concerns. Throughout the course, Pat came to see understanding GenAI as a necessary future skill, claiming that students should be allowed to use it because it is a "\textit{tool that people are probably going to use in the future, and it's better to understand it than not}". Pat perceived GenAI as a tool for personal empowerment, one that expands his abilities to be "better and faster". This view aligned with his belief that society globally values progress through efficiency and quality. Consequently, he concluded that GenAI would be celebrated as an instrument for achieving higher standards.
\subsubsection{Change in students' perceptions of GenAI’s impact on their careers over time}

In the first interview, Pat held a consistent view of GenAI's role in the professional world. He believed that "\textit{AI is definitely gonna have its place in the job market}" and that since "\textit{everyone will use AI whether they like it or not}", people would have to "adapt or die". He was not concerned about its impact on his own entrepreneurial career, confidently stating he does not "\textit{think that the thing that AI is good at is something that I'm good at, and vice versa}". Pat was "super pumped" for GenAI's integration into daily life, viewing it as a tool that would improve efficiency.

In the fourth interview, this view did not change, even as Pat began using generative GenAI to complete his schoolwork. In the final interview, when asked if his understanding of GenAI had impacted his career choices or if he was concerned about its effects, his answer was a consistent and repeated "No". His rationale remained unchanged: he saw GenAI’s future role in his work as simply making tasks "easier". Pat believed that GenAI use in the workforce will be universal, and will be seamlessly integrated into new technology that people may not even realize that the technology they are using utilizes GenAI. He maintained that GenAI will become a necessity to use to stay competitive in the workforce. 
Although Pat used GenAI extensively for academic tasks that he devalued, he never explored its potential uses for his actual passion, business, or entrepreneurship. Pat's conviction that GenAI could not perform the same tasks he excels at remained untested, as Pat never attempted to leverage it for his business-related activities. 
Pat stayed consistent in his attitude that GenAI will not affect his chosen career negatively, and will be ubiquitously integrated into future technology that is used in jobs and daily life. Instead of a threat, GenAI will help his future career path.

\subsubsection{Change in students' expectations about their own career outcomes over time}

In the first interview, Pat's primary objective was to graduate from his current university, mainly to "make my parents happy" and "get done with doing school," rather than for specific career goals. He is not satisfied with the jobs offered by his current business management degree, finding them "terrible" and paying less than his employees from his painting business. Pat's main concern was "\textit{being able to afford my degree and being able to continue paying for it}", as he paid for college himself. He had enjoyed running through a collegiate entrepreneurship program painting business over the summer, but as of interview 1, became "bored" of it and wished to "start a new, different, venture" but had no clear idea of what that would be. He planned to secure "the highest paying job" around his current town, and "try to figure out what's next" after graduation. When asked if Pat had concerns about GenAI affecting his career expectancy, he stated "No," explaining that GenAI is good at creative tasks that he is not, and that people would simply have to "adapt or die" to the new technology without it threatening his specific path.

Despite significant changes in his life and academic habits, Pat's career outcome expectancy remained unchanged. When asked if his new understanding of GenAI had impacted his career choices or if he was concerned, Pat's answer was a consistent "no". His narrative about his career had shifted to focus on his immediate professional circumstance. He described his current employment at a gas station as a "shitty job" that "pays the bills". 
Throughout the span of the course, Pat's career outcome expectancy remained consistently unaffected. Increased GenAI proficiency, a potential career skill, did not alter his belief about his ability to get a job in his desired field. He did not connect the changes in his immediate job situation or his increased use of GenAI for schoolwork to his long-term career prospects. Instead, he consistently and directly reported that GenAI was irrelevant to his entrepreneurial goals. Therefore, based on his repeated statements, we concluded that there was no change in his perception of how GenAI would affect his future career. However, he says he has not used GenAI at all outside of the school environment.

\subsection{Jay - Raised ethical skills, Impeded AI skills}  

Jay is an aspiring author, pursuing an English major and a Spanish minor. He started the semester with confidence in his ability to outperform GenAI, using it to show the importance of human emotion in storytelling. Jay's initial confidence in his career expectations decreased over the semester as his perception of GenAI's role shifted. His awareness of GenAI's rapid adoption in creative industries led to anxiety surrounding his career outcome expectancy, making him "nervous to do any type of writing professionally". Jay's experiences show a student whose professional confidence was negatively impacted not by their own use of GenAI, but by their perception of its threat to the value of human connection in their field. 

\subsubsection{Change in effects of using GenAI over time}

In the first interview, Jay described his first experience with AI-related technology as an older child when he saw a human-interest video on YouTube. He concluded that AI doesn’t capture the “nuances of human emotion.” He initially used GenAI to bypass paywalls on book summary websites to get the "big picture of this chapter" and tested its limits for his English classes, concluding that GenAI "cannot" succeed at an undergraduate level. He agreed with other peers in the English department who also viewed the output as sub-par quality. Jay argued that GenAI's simple writing emphasizes "\textit{how important it is to really understand the emotional well being of the character}" in human writing. 
Jay also expressed significant apprehension about the ethical consequences and potential abuse of generative AI. His primary concern involves the malicious application of GenAI as a tool to spread misinformation, particularly within political campaigns. Furthermore, as an artist, Jay is deeply troubled by the unauthorized use of artwork for training GenAI models without the creators' consent or acknowledgment, an issue that resonates with his professional identity.
Despite his concerns, he valued GenAI's role as a tool for educational accessibility. He found it gratifying that GenAI could circumvent paywalls, correct grammar, and act as a language tutor. Therefore, Jay believed GenAI could provide access to resources that were previously limited by logistical barriers such as time and money.

In the fourth interview, Jay was minimally using GenAI, guided by a new self-imposed limit of 10 minutes daily due to concerns about GenAI's environmental impact created during a class ethics activity on making your own GenAI use guidelines \cite{landesmanIntegratingPhilosophyTeaching2024}. His practical use of GenAI was restricted to brief tasks, such as changing verb tenses to avoid paying for other grammar-correcting software. Jay found that GenAI's suggested edits often contradicted with his intended meaning, leading him to conclude, "\textit{I actually like what I wrote more than what this AI, seems to think I should be writing}". This experience boosted his confidence as a writer, making him realize that, "\textit{I think I'm better at writing than what I assumed that I was}". 
In addition, Jay questioned GenAI's future role in societal narratives, speculating on a potential socioeconomic dichotomy: would free GenAI be perceived as an inferior tool for the lower classes, or would it become a celebrated symbol of status? This uncertainty extended to whether GenAI proficiency would eventually become a mandatory expectation for full integration into modern society.
 
Jay’s perspective on GenAI is one of pronounced and evolving tension, culminating in the view that its detriments ultimately outweigh its benefits. Initially, his skepticism was rooted in his initial ethical perceptions based on observed output, but it deepened over the semester into a profound critique based on multifaceted impacts. This shift was driven by a growing awareness of GenAI's environmental costs and the uncompensated use of artists' work for training data. As these ethical and professional anxieties intensified, Jay’s interactions with GenAI paradoxically strengthened confidence in his own capabilities by declining to use GenAI. This occurred particularly after observing the technology’s inability to meaningfully improve his own writing.
This evolving attitude led to tangible changes in Jay's behavior and outlook. He enacted self-imposed usage limits that, while ethically motivated, also impeded further development of his own GenAI skills due to a lack of use. His objection to GenAI's role in creative fields became more defined, leading him to intentionally avoid its use in his artistic practice. However, his final position is not one of total rejection. In reflection, Jay now acknowledges GenAI’s value as a tool for the “average Joe,” recognizing its potential to help non-specialists learn new skills and engage in creative expression, even as he maintains a firm stance against its use in professional artistic contexts.

\subsubsection{Change in students' perceptions of GenAI’s impact on their careers over time}

In the first interview, Jay was "\textit{really worried that it will take jobs away from people, especially creative jobs}", citing stories of copywriters being laid off. This concern fueled him to become more passionate about his writing, stating, "\textit{my job cannot be taken by computers. I refuse, I refuse to let that happen}". While he believed GenAI should be used for "tedious jobs" rather than creative ones, he also saw a use for it in his own career, showing interest in learning how GenAI could affect web design to market himself as an author. He speculates that GenAI is currently a popular trend and hopes its application will become more specialized over time, ultimately limited to areas where it is genuinely beneficial and not "detrimental" to creative or valued human-centered jobs.

In the fourth interview, Jay's sense of whether he will have the agency to choose to use GenAI in his future career became context-dependent. He believed that, "\textit{I won't have a choice if I'm part of a business setting that requires me to use it}". He would consider voluntarily using it if he had the choice on personal writing projects, but it would be dependent on how much his "employer expected from me." He was critical of businesses using GenAI for creative tasks when he could afford to pay a human to do the work. His concern was that the rush to adopt GenAI would cause people to "\textit{completely forget the fact that there's like, a human to human relationship when it comes to reading}".

Jay's sense of agency evolved significantly. After initially refusing to be replaced by technology, he developed a more pragmatic view by the final interview. He acknowledged he would have little agency in his choice if a future business required him to use GenAI, but would retain the agency to choose if he was completely in charge of his own writing projects. The fear of copywriters being replaced became a direct reason why he was now "nervous to do" that type of work, altering his professional path.

\subsubsection{Change in students' expectations about their own career outcomes over time}

In the first interview, Jay expressed high certainty in his career outcome expectations as a writer, specifically hoping to work for a nonprofit. While he was uncertain about the specific "capacity or to what degree" he would work in that field, his belief in his writing abilities was prominent. Jay's response to the threat of GenAI was to strengthen his skills, viewing GenAI as a flawed competitor he could beat.

In the fourth interview, his career outcome expectancy had declined. The specific goal of working in a nonprofit was not mentioned; instead, his focus shifted to a fear of entering the writing profession. Jay stated that he is "nervous to do any type of writing professionally at this point" due to the quick adoption of GenAI in writing fields. Jay reflected that his future career search will be guided by intrinsic fulfillment rather than external judgment of others' expectations. Citing loyalty and integrity as core personal values, he stated an intention to seek a profession that not only reflects this ideology but also allows him to positively contribute to his community.

Between the first and fourth interviews, the change in Jay's career outcome expectancy was negative. His initial goal of working as a writer in the nonprofit sector was replaced by anxiety about the viability of any professional writing career. The technology he initially saw as an incapable competitor transitioned into what made him question the possibility of pursuing a professional writing career altogether.
In conclusion, Jay shifted from pursuing a specific career target to a broader search for a profession that offers personal fulfillment and aligns with his own judgment.

\section{Limitations}
This study is based on two in-depth cases and is not meant to be statistically generalizable. Instead, it offers rich insights into how GenAI affects student learning, ethics, and career thinking in a creative media course with some programming. These dynamics may differ in more technical computing contexts. While we tracked changes across one semester, we could not assess long-term outcomes. Self-reported data may contain recall or desirability bias, though triangulation and comparison between statements made in the first and fourth interview helped mitigate this.

\section{Discussion}

Our findings demonstrate that designing for learning with generative AI exemplifies what Rittel and Webber termed a "wicked problem"—a challenge that lacks clear solutions, involves competing values, and generates new problems when addressed \cite{RittelDilemmasTheory1973}. The cases of Pat and Jay illustrate how interventions aimed at improving one educational outcome (e.g., GenAI skills or ethics) can inadvertently undermine others, producing the entangled dynamics described in wicked problem literature \cite{bass_coda_2022, roberts_wicked_2000}.

While some studies have examined GenAI's effect on skill acquisition \cite{chan_deconstructing_2023}, anxiety \cite{yasar_effect_2025}, or career outlooks \cite{sallam_anxiety_2024}, few explore how these outcomes interact over time. Our longitudinal analysis reveals how students actively negotiate and revise their ethical and career stances in response to new experiences with GenAI, often inconsistently or paradoxically. Jay’s ethical ambivalence, for example, aligns with wicked problem theory’s emphasis on unresolved tensions and shifting judgments.

Our findings raise questions about assumptions in Social Cognitive Career Theory (SCCT), particularly regarding the formation of self-efficacy. SCCT posits that self-efficacy grows when individuals attribute success to their own effort and skills \cite{lentSocialCognitiveModel2013}. Yet Pat’s case challenges this notion. He frequently used GenAI to “get all the right answers,” fully aware he was learning less, yet still expressed a sense of academic confidence. This disconnect echoes findings on the illusion of competence in technology-assisted learning \cite{prather2024widening}, but extends them by showing how self-efficacy can persist even with explicit acknowledgment of GenAI dependency. Our theoretical conceptions of self-efficacy may need to be enriched, given the existence of technologies with GenAI's actual or perceived capabilities. 

Jay, by contrast, used GenAI less due to increased concerns about environmental impact and data sourcing ethics, which impeded his GenAI use skill development. Notably, these concerns selectively influenced his behavior: he avoided GenAI to uphold artistic consent but rationalized bypassing academic paywalls using the same tools. This contradiction—valuing ethical use in one domain but justifying unauthorized use in another—illustrates the complex and often inconsistent moral reasoning students navigate when using GenAI.

Pat’s and Jay’s cases also reveal a broader misalignment between education and perceived and evolving workplace realities. Despite course modules on GenAI’s professional applications, students struggled to project how GenAI might affect their own careers. This disconnect exhibits how particularly with GenAI students are forming career expectations often not from curricula but from external narratives, such as news headlines, peer assumptions, and cultural framing, echoing findings from recent survey work \cite{sallam_anxiety_2024, hwang_influence_2024}.

We argue that current educational systems are ill-equipped to help students develop accurate mental models of GenAI’s role in the labor market. This leaves students susceptible to anxiety, detachment, or premature resignation. Designing better “future-of-work” literacy may require more active scaffolding: simulations, role-playing as managers vs. workers, and integration of career counseling with ethics education.

Our findings point to a few key takeaways for instructors working with GenAI in the classroom. It is important to measure more than just how well students perform on tasks, as focusing only on outcomes can hide drops in ethics or career confidence \cite{kolko_wicked_2012, mishra_technological_2007}. Students also need help thinking ahead to how GenAI shapes the job market, not just how to use the tools. That means supporting them in understanding automation and labor trends, while also integrating justice-centered approaches positioning students as having agency in living out their values and creating counter-narratives, such as in Eversen et al's work \cite{everson_key_2022}.

Future research should build multi-dimensional measurement tools for factors such as critical and ethical GenAI use self-efficacy. Longitudinal studies should trace how student perspectives evolve when given different ethical, career, and disciplinary framings. We need curricular and course designs that also foster student identity, long-term motivation, and ethical reasoning across diverse contexts and potential careers and visions for the future. This includes patterns and strategies for integrating GenAI, and reflecting, discussing, and jointly co-constructing and co-designing GenAI integration and use in education.

\begin{acks}

[Anonymized for review. This will include brief acknowledgments to collaborators, participants, and one funding source.]

\end{acks}

\balance 

\clearpage
\bibliographystyle{acm}
\bibliography{references}
\end{document}